\documentclass[%
 reprint,
 superscriptaddress,
 preprintnumbers,
 amsmath,amssymb,
 aps,
floatfix,
]{revtex4-2}

\usepackage[utf8]{inputenc}
\usepackage[T1]{fontenc} 
\usepackage{graphicx}
\usepackage{dcolumn}
\usepackage{bm}
\usepackage{xcolor}
\usepackage{hhline}

\renewcommand{\arraystretch}{1.2}
\makeatletter
\newcommand{\mathleft}{\@fleqntrue\@mathmargin0pt}
\newcommand{\mathcenter}{\@fleqnfalse}
\makeatother

\begin{document}

\preprint{ADP-23-11/11220}
\preprint{DESY-23-049}
\preprint{LTH 1337}

\title{Weak decay constants of the neutral pseudoscalar mesons from lattice QCD+QED}

\author{Z.R.~Kordov}
\affiliation{%
 CSSM, Department of Physics, University of Adelaide, SA, Australia
}%
\author{R.~Horsley}
\affiliation{School of Physics and Astronomy, University of Edinburgh, Edinburgh EH9 3FD, UK}
\author{W.~Kamleh}%
\affiliation{%
 CSSM, Department of Physics, University of Adelaide, SA, Australia
}%
\author{Y.~Nakamura}
\affiliation{RIKEN Center for Computational Science, Kobe, Hyogo 650-0047, Japan}
\author{H.~Perlt}
\affiliation{Institut für Theoretische Physik, Universit\"at Leipzig, 04109 Leipzig, Germany}
\author{P.E.L.~Rakow}
\affiliation{Theoretical Physics Division, Department of Mathematical Sciences,University of Liverpool, Liverpool L69 3BX, UK}
\author{G.~Schierholz}
\affiliation{Deutsches Elektronen-Synchrotron DESY, Notkestr. 85, 22607 Hamburg, Germany}
\author{H.~St\"uben}
\affiliation{Regionales Rechenzentrum, Universit\"at Hamburg, 20146 Hamburg, Germany}
\author{R.D.~Young}%
\affiliation{%
 CSSM, Department of Physics, University of Adelaide, SA, Australia
}%
\author{J.M.~Zanotti}
\affiliation{%
 CSSM, Department of Physics, University of Adelaide, SA, Australia
}%

\collaboration{CSSM/QCDSF/UKQCD Collaboration}
\noaffiliation

\date{\today}

\begin{abstract}
With increasing requirements for greater precision, it becomes essential to describe the effects of isospin breaking induced by both quark masses and electro-magnetic effects. In this work we  perform a lattice analysis of the weak decay constants of the neutral pseudoscalar mesons including such isospin breaking effects, with particular consideration being given to the state mixing of the $\pi^0$, $\eta$ and $\eta^\prime$. We also detail extensions to the non-perturbative RI$^\prime$-MOM renormalization scheme for application to non-degenerate flavour-neutral operators which are permitted to mix, and present initial results. Using flavour-breaking expansions in terms of quark masses and charges we determine the leptonic decay constants for the $\pi^0$ and $\eta$ mesons,
demonstrating in principle how precision determinations of all neutral pseudoscalar decay constants could be reached in lattice QCD with QED and strong isospin-breaking accounted for.
\end{abstract}

\maketitle

The pseudoscalar mesons are of particular interest in QCD due to their relationship to the spontaneous chiral symmetry breaking (SCSB) of the vacuum. In the chiral limit and without electromagnetism, the pseudoscalar octet species are understood to be massless Nambu-Goldstone bosons of SCSB, whilst the flavour-singlet $\eta^\prime$ acquires a mass through an Adler-Bell-Jackiw anomaly \cite{Bell:1969ts,Adler:1969gk,Adler:1969er} due to non-conservation of the singlet axial-vector current upon quantization. This anomaly is simultaneously understood to arise from the non-trivial topological structure of the QCD vacuum \cite{PhysRevLett.37.8}, and is often studied in large-$N_c$ perturbation theory \cite{Witten:1978bc,Veneziano:1979ec} whereby chiral symmetry may be restored. Moreover, a similar ${U}_A(1)$ anomaly in QED further breaks chiral symmetry and allows unique electroweak decays of the flavour-neutral (FN) pseudoscalar (PS) mesons. Hence, with their relationship to SCSB and the dynamics of the physical vacuum, the FN PS mesons and their decays offer unique tests of fundamental physics which may prove fruitful in the search for physics beyond the standard model \cite{Gan:2020aco}, for which experimental searches using them have recently been proposed \cite{REDTOP:2022slw}.

There are a variety of existing studies of the flavour-neutral pseudoscalar meson decay constants using various approaches \cite{Gan:2020aco}, however only lattice QCD facilitates the non-perturbative calculation of decay constants from first principles. 
While some attempts have been made to determine the isospin-breaking corrections to the charged pseudoscalar meson decay constants, including QED effects (e.g. \cite{Boyle:2022lsi}), to date all of the available lattice studies (e.g. \cite{Dudek:2013yja,Ottnad:2017bjt,Bali:2021qem}) in the flavour-neutral sector have been performed using degenerate up and down quark masses, and without QED, and hence the $\pi^0$ considered therein is the familiar isovector state.
In this study, as in our previous investigation of the state mixing \cite{CSSMQCDSFUKQCD:2021rvs,Kordov:2019oer}, we include the effects of strong isospin-breaking and QED, with a result being non-trivial state mixing between the $\pi^0$ and $\eta$/$\eta^\prime$. Rather than incorporating QED effects perturbatively, we generate background QED fields together with the QCD fields  \cite{Horsley:2015eaa,Horsley:2015vla}. Since our focus is on the FN PS mesons, we do not calculate decay constants of the charged PS meson, and hence we avoid the delicate cancellation that would be required between photons in the final state and and an infra-red divergence \cite{DiCarlo:2019thl,Christ:2023lcc}. In addition to the FN $\pi^0,\, \eta,\, \eta^\prime$ we also include the $K^0$ primarily to help constrain our quark charge/mass extrapolations which are introduced in Section~\ref{sec:fbexpansions}.

Our primary aim is the lattice determination of weak decay constants of FN PS meson states, which we choose to define through the Euclidean time component of the axial-vector current,
\begin{equation*}
    \langle \Omega | \, A_4^f \, |  n(\vec{p}=0) \rangle \, = \, M_n  F_n^f, \quad A_4^f = \bar{q}_{f}\gamma_4\gamma_5 q_{f},
\end{equation*}
\begin{equation}
    f=u, \, d, \, s, \quad n = \pi^0, \, \eta, \, \eta^\prime, \label{eqn:QFBdecayconstdef}
\end{equation}
in this case with respect to axial-vector currents in the quark-flavour basis instead of the octet-singlet basis, and where $M_n$ denotes the mass of the state $n$. We also determine the $K^0$ decay constant defined similarly via
\begin{equation}
    \left\langle \Omega | \, \bar{s}\gamma_4 \gamma_5 d \, | K^0(p) \right\rangle \, = \, M_{K^0} \, F_{K^0}.
\end{equation}
The definitions for the FN and $K^0$ decay constants are distinct in that the FN axial-vector currents can couple non-trivially to each of the three FN PS meson eigenstates. In contrast, the $K^0$ has a trivial flavour content and therefore only couples through one flavour channel.

In this work we do not parametrize the flavour-neutral decay constants by mixing angles as is common in the isospin-symmetric literature \cite{Gan:2020aco}, as this process becomes unnecessarily complicated when considering three states. In addition to the lattice decay constants we discuss renormalization in the non-perturbative RI$^\prime$-MOM scheme and make determinations of the renormalization factors for each of our axial-vector operators, in particular developing the process for the inclusion of QED corrections. We also fit quark mass and charge parametrizations to our renormalized decay constants in order to present results at the physical point. These developments, in concert with continuum and infinite volume extrapolations which are left for future work, facilitate in principle the precision calculation of all neutral PS meson decay constants on the lattice. 

\section{Lattice calculation of decay constants}

To extract the flavour-neutral decay constants we construct the lattice correlation functions
\begin{equation}
C^{AP}_{f f^\prime, \, l}(t) = \frac{1}{N_t}\sum_{t^\prime}\sum_{\vec{x}, \vec{y}} \langle \Omega | \, \mathcal{A}_{f}(\vec{y},t+t^\prime) \, \mathcal{P}^{(l)}_{f^\prime}(\vec{x},t^\prime)^\dagger \, | \Omega \rangle, \label{eqn:APcorrelators}
\end{equation}
where the operators present are $\mathcal{A}_{f}\equiv A^{f}_4=\bar{q}_{f}\gamma_4\gamma_5 q_{f}$ and $\mathcal{P}^{(l)}_{f^\prime}=\bar{q}^{(l)}_{f}\gamma_5 q_{f}^{(l)}$ for quark flavours $f,f^\prime=u,d,s$, and the quark-field superscript $(l)$ 
denotes one of
two levels of gauge-covariant Gaussian smearing applied at the source. The number of sites in the time extent of the lattice is denoted $N_t$, and we have included an average over 
independent source locations
to improve the signals of disconnected contributions. The correlation function label $AP$ simply indicates the types of operators present: axial-vector and pseudoscalar. This set of correlation functions may be interpreted as a $3\times 6$ matrix of correlation functions, $C^{AP}_{fj}(t)$, with the row index $f$ enumerating three flavours of axial-vector operator $\mathcal{A}_{f}$, and the column index $j$ enumerating three flavours ($f'$) of pseudoscalar operator $\mathcal{P}^{(l)}_{f^\prime}$ for each of two smearing levels ($l$). Note that the axial-vector remains un-smeared since the decay constants are defined using local operators.

In order to isolate distinct eigenstates on the lattice we
compute the pseudoscalar correlation functions
\begin{equation}
C^{PP}_{f f^\prime, \, l \, l^\prime}(t) = \frac{1}{N_t}\sum_t\sum_{\vec{x}, \vec{y}} \langle \Omega | \, \mathcal{P}^{(l)}_{f}(\vec{y},t+t^\prime) \, \mathcal{P}^{(l^\prime)}_{f^\prime}(\vec{x},t^\prime)^\dagger \, | \Omega \rangle, 
\label{eqn:PPcorrelators}
\end{equation}
where each $f,f'$ and $l,l'$ index again enumerates the three light quark-flavours and two levels of quark-smearing, respectively. These pseudoscalar correlation functions may be interpreted as a $6\times 6$ matrix, $C^{PP}_{ij}(t)$, with indices $i$ and $j$ enumerating three flavours of pseudoscalar operator $\mathcal{P}^{(l)}_{f^\prime}$ for each of two smearing levels. For this study we note that one of the two smearing levels used is trivial (i.e. not smeared). 

In each case where flavour-neutral meson operators are utilized, the resulting correlation functions receive disconnected quark-loop contributions, and we approximate these contributions using diluted $\mathbb{Z}_2$ noise-sources as detailed in our previous work \cite{CSSMQCDSFUKQCD:2021rvs}. The motivation for including additional operators with varied levels of smearing is to better distinguish the $\eta'$ from nearby states such as the resonant $\eta(1295)$ and $\pi(1300)$ or potential multi-particle levels. Since the computationally expensive disconnected contributions may be smeared at source and sink after inversion, extending the variational basis in this way incurs little additional cost.

The spectral decomposition for our pseudoscalar correlation functions in the limit of very large Euclidean time with relativistic normalization is approximated by
\begin{equation}
	C^{PP}_{ij}(t) \, \xrightarrow[t \rightarrow \infty] \, \sum_{n=1}^6 \frac{L^3}{2 M_n } \langle \Omega | \mathcal{P}_{i} | n \rangle \, \langle n | \mathcal{P}^\dagger_{j} | \Omega \rangle \, e^{-M_n t}, 
\end{equation}
with only the six least-energetic states contributing, and $\mathcal{P}_i$ one of the six $\mathcal{P}^{(l)}_f$ described above. In this limit there exist eigenvectors $\vec{v}_m$ of the matrix of correlation functions which satisfy
\begin{equation}
    \sum^6_{j=1} \, \langle \Omega | \mathcal{P}_j | n \rangle \, [\vec{v}_m]_j = \delta_{nm} \sqrt{\sum^6_{i=1} \, |\langle \Omega | \mathcal{P}_i | n \rangle|^2}, \label{eqn:evecdelta}
\end{equation}
and hence allowing the diagonalization of the (real and symmetric) matrix $C^{PP}_{ij}(t)$ via
\begin{equation}
	\vec{v}_n^T \, C^{PP}(t) \, \vec{v}_n =  \frac{L^3}{2 M_n} \, e^{-M_n t} \sum^6_{j=1} |\langle \Omega | {\mathcal{P}}_{j} | n \rangle |^2. \label{eqn:PPdiag}
\end{equation}
Of course, for pseudoscalar mesons on a lattice with finite time extent $T$ and periodic boundary conditions, the full time dependence of the two-point functions amounts to the replacement $e^{-M_n t} \rightarrow (e^{-M_{n} \, t} \, + \, e^{-(T-t)M_{n}})$, although we leave this replacement implicit in our current presentation for simplicity.

We determine the eigenvectors $\vec{v}_n$ from the generalized eigenvalue problem (GEVP), 
see e.g.~Refs~\cite{Luscher:1990ck,Michael:1982gb,Blossier:2009kd,Mahbub:2009nr},
\begin{equation}
	C^{PP}(t_0)^{-1} \, C^{PP}(t_0+\delta t) \, \vec{v}_n = e^{-M_n\delta t} \, \vec{v}_n, \label{eqn:ppGEVP}
\end{equation}
for $\delta t = 1$, and the largest $t_0$ which results in reasonable numerical stability, which we find to be $t_0=4$.

The spectral decomposition of Equation~\eqref{eqn:APcorrelators} at large Euclidean time yields the expression
\begin{equation}
	C^{AP}_{fj}(t) \, \xrightarrow[t \rightarrow \infty] \, \sum_{n=1}^6 \frac{L^3}{2 M_n } \langle \Omega | \mathcal{A}_{f} | n \rangle \, \langle n | \mathcal{P}^\dagger_{j} | \Omega \rangle \, e^{-M_n t}, \label{eqn:APspecdecomp}
\end{equation}
and utilizing Equations~\eqref{eqn:evecdelta} and \eqref{eqn:PPdiag} we may compute the desired un-renormalized decay constant of Equation~\eqref{eqn:QFBdecayconstdef}, $F^f_n = \langle 0 | \mathcal{A}_{f} | n \rangle / M_n$, at large Euclidean time as
\begin{equation}
F^f_n = \sqrt{\frac{2}{ M_n L^3}} \, e^{M_{n} \, t} \, \frac{\sum^6_{j} C^{AP}_{f j}(t) \, [\vec{v}_n]_{j}}{\sqrt{\vec{v}_n^T \, C^{PP}(t) \, \vec{v}_n \, e^{M_{n} \, t}}}. \label{eqn:rawlatticeFNdecayconst}
\end{equation}

The lattice calculation of the $K^0$ decay constant is relatively straightforward. The correlation functions needed are
\begin{equation}
	C^{AP}_{K^0}(t) = \sum_{\vec{x}, \vec{y}} \langle \mathcal{A}_{K^0}(\vec{y},t) \, \mathcal{P}^\dagger_{K^0}(\vec{x},0) \rangle,
\end{equation}
\begin{equation}
	 C^{PP}_{K^0}(t) = \sum_{\vec{x}, \vec{y}} \langle \mathcal{P}_{K^0}(\vec{y},t) \, \mathcal{P}^\dagger_{K^0}(\vec{x},0) \rangle,
\end{equation}
where
\begin{equation}
	\mathcal{A}_{K^0}=\bar{s}\gamma_4 \gamma_5 d, \quad \mathcal{P}_{K^0}=\bar{s}\gamma_5 {d}. \label{eqn:K0operators}
\end{equation}
The decay constant calculation then proceeds in the limit of large Euclidean time as
\begin{equation}
	F_{K^0} = \sqrt{\frac{2}{ M_{K^0} L^3}} \, e^{M_{K^0}  t} \, \frac{ C^{AP}_{K^0}(t)}{ \sqrt{C_{K^0}^{PP}(t) \, e^{M_{K^0}  t} }}. \label{eqn:rawpi+decayconst}
\end{equation}

\section{Operator renormalization}

In order to make sensible comparisons between our lattice decay constant results and other determinations, we must renormalize our interpolating operators to remove their dependence on the lattice regulator. In the present study we utilize the non-perturbative RI$^\prime$-MOM renormalization scheme, taking care to detail the application to quark-flavour operators with non-degenerate electric charges. 

\subsection{$K^0$ operator}

We will first detail the RI$^\prime$-MOM procedure for calculating the renormalization Z-factors relevant to the axial-vector interpolating operator used for the $K^0$, which closely resembles existing presentations in the literature \cite{Martinelli:1994ty}. The renormalized axial-vector operator is defined multiplicatively as
\begin{equation}
	\tilde{\mathcal{A}}_{K^0}(\mu) = Z^{K^0}_{A}\left( \mu a \right)\mathcal{A}_{K^0}(a), \label{eqn:renormoperator}
\end{equation}
where the Z-factor is labelled to denote that it is applied to an axial-vector ($A$) operator having the quark-flavour content of a $K^0$. The resulting operator $\tilde{\mathcal{A}}_{K^0}(\mu)$ is independent of the lattice regularisation scheme. Now suppressing the lattice spacing $a$, the renormalization condition in RI$^\prime$-MOM makes use of the amputated vertex function
\begin{equation}
	\Gamma_{K^0}^{A}(p) = S_{s}^{-1}(p)G_{K^0}^{A}(p)S_{d}^{-1}(p),
\end{equation}
where the above Green's function is given by
\begin{equation}
		G_{K^0}^{A}(p) = \frac{1}{V_4}\sum_{x,y,z} \, e^{-ip(x-y)} \, \big\langle s(x) \, \mathcal{A}_{K^0}(z) \, \bar{d}(y) \big\rangle ,
\end{equation}
where $s(x)$ and $d(x)$ are strange and down quark fields respectively. The momentum-space propagator of an $f$-flavoured quark is given by
\begin{equation}
	S_{f}(p) = \frac{1}{V_4} \sum_{x,y} \, e^{-ip(y-x)} \, S_{f}(y;x),
\end{equation}
which is in practice calculated using a momentum source \cite{Gockeler:1998ye}, and we have denoted the four-volume of the lattice with $V_4$.

We define the renormalized amputated vertex function using Equation~\eqref{eqn:renormoperator} as
\begin{equation}
    \tilde{\Gamma}_{K^0}^{A}(p) = \frac{1}{\sqrt{Z_{q_s}Z_{q_{d}}}}  \, Z^{K^0}_{A} \, \Gamma_{K^0}^{A}(p),
\end{equation}
where the $f$-flavoured quark field renormalization factor $Z_{q_f}$ is given by
\begin{equation}
	Z_{q_f}(p) = \frac{\textrm{Tr}\left[ -i\sum_{\lambda} \, \gamma_\lambda \, \sin(p_\lambda) \, S_f^{-1}(p) \right]}{ 12 \sum_\rho \, \sin^2(p_\rho )}.
\end{equation}
We now impose the condition that the renormalized amputated vertex function be equal to its tree-level value in perturbation theory, $\Gamma^{Born}_{A_\mu} = \gamma_5 \gamma_\mu$, at the scale $p^2 = \mu^2$. Hence the required Z-factor is calculated as
\begin{equation}
	\frac{1}{Z^{K^0}_{A}}=\frac{1}{12\sqrt{Z_{q_{s}}Z_{q_{d}}}}\textrm{Tr}\left[\Gamma^{K^0}_{A}(p) (\Gamma_{A_\mu}^{Born})^{-1}\right]\Big|_{p^2=\mu^2}.
\end{equation}
In practice, the operator renormalization factors are calculated at a range of scales (momenta), as for sufficiently large momenta the flavoured axial-vector operator renormalization is independent of scale. Moreover, it is known that there exist discretization errors in the RI$^\prime$-MOM renormalization factors beginning at $(ap)^2$, and so the Z-factors are extracted by first identifying the scale-independent region and then linearly extrapolating to $(ap)^2=0$ in order to minimize these effects. We will see in our results that considering only the leading-order discretization errors, $\propto (ap)^2$, gives a very good description of the data.

\subsection{Flavour-neutral operators}

\begin{figure}
	\centering
	\includegraphics[width=0.45\textwidth]{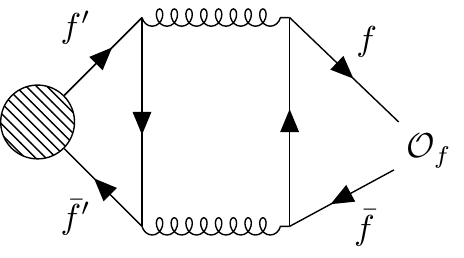}
	\caption{A leading-order, $\mathcal{O}(g^4)$, diagram contributing to the operator mixing, where $\mathcal{O}_f$ denotes our quark-flavour operator of flavour $f$, and the lined circle an external state.} \label{fig:operatormixing}
\end{figure}

In the case of flavour-neutral quark-bilinears, we must allow for operator mixing which can occur through diagrams such as that illustrated in Figure~\ref{fig:operatormixing}. While in principle we may also have gluonic operators contributing to the renormalized quark-flavour operator, for the present study we will only consider the mixing of quark operators, expressed as
\begin{equation}
	\tilde{\mathcal{A}}_f(\mu) = \sum_{f'} Z_A^{f f'}\left( \mu a \right)\mathcal{A}_{f'}(a). \label{eqn:FNrenormoperator}
\end{equation}
In this expression the $f$,$f'$ are indices possessed by the Z-factor, which may hence be interpreted as a matrix in flavour which acts to transform the lattice operators to the new renormalization scheme,
\begin{equation}
	\begin{bmatrix} 
		\tilde{\mathcal{A}}_u(\mu)\\
		\tilde{\mathcal{A}}_d(\mu)\\
		\tilde{\mathcal{A}}_s(\mu)\\
	\end{bmatrix} 
	=
	\begin{bmatrix} 
		Z^{uu}_A & Z^{ud}_A & Z^{us}_A \\
		Z^{du}_A & Z^{dd}_A & Z^{ds}_A \\
		Z^{su}_A & Z^{sd}_A & Z^{ss}_A \\
	\end{bmatrix} 
	\begin{bmatrix} 
		\mathcal{A}_u(a)\\
		\mathcal{A}_d(a)\\
		\mathcal{A}_s(a)\\
	\end{bmatrix} . \label{eqn:matrixZs}
\end{equation}

As before we begin with the non-perturbative calculation of a bare, amputated vertex function
\begin{equation}
	\Gamma^A_{f f'}(p) = S_{f'}^{-1}(p) \, G^A_{f f'}(p) \, S_{f'}^{-1}(p),
\end{equation}
with the Green's function
\mathleft
\begin{equation*}
	G^A_{f f'}(p) = \frac{1}{V_4}\sum_{x,y,z} \, e^{-ip(x-y)} \, \langle q_{f'}(x) \, \mathcal{A}_{f}(z) \, \bar{q}_{f'}(y) \rangle 
\end{equation*}
\mathcenter
\begin{equation*}
    = \frac{1}{V_4}\sum_{x,y,z} \, e^{-ip(x-y)} \, \bigg( \delta_{f f^{\prime}} \, S_{f}(x;z) \, \gamma_\mu \gamma_5 \, S_{f}(z;y) \, 
\end{equation*}
\begin{equation}
    -  \, S_{f^\prime}(x;y) \, \textrm{Tr}\left[ S_{f}(z;z) \gamma_\mu \gamma_5 \right] \bigg). \label{eqn:3ptGreenfunc}
\end{equation}
Notable in the above expression is the second term, present for all combinations of $f$ and $f^\prime$, which corresponds to the operator of interest being inserted on a closed quark loop and facilitates the operator mixing numerically. The renormalized vertex function is obtained by an application of Equation~\eqref{eqn:FNrenormoperator}, yielding
\begin{equation}
	\tilde{\Gamma}^A_{f f'}(p) = \frac{1}{Z_{q_{f'}}}\sum_{g} Z_A^{f g}\Gamma^A_{g f'}(p). \label{eqn:renormampvertex}
\end{equation}
Enforcing the renormalization condition from tree-level perturbation theory, which for our matrix of vertex functions has the form
\begin{equation}
	\tilde{\Gamma}^A_{f f'}(p) = \delta_{f f'} \Gamma_{A_\mu}^{Born},  \label{eqn:FNbornterm}
\end{equation}
we find 
\begin{equation}
	\left(Z^{-1}_A\right)^{f f'} = \frac{1}{12 Z_{q_{f'}}(p)} \textrm{Tr}\left[\Gamma^A_{f f'}(p) (\Gamma_{A_\mu}^{Born})^{-1}\right]\Big|_{p^2=\mu^2}. \label{eqn:inverseZmatrix}
\end{equation}
The result above, $\left(Z^{-1}_A\right)^{f f'}$, is the $(f,f')$ component of the matrix-inverse of the Z-factor matrix as it is given in Equation~\eqref{eqn:matrixZs}. 

In contrast to the $K^0$ operator, the anomalous dimension of the FN operators results in a scale dependence that may be observed even at large momenta. However for the range of scales (momenta) we consider, and at our current level of precision, we will see that no scale dependence can be observed. 

\section{Results and analysis}

The lattice results in this study have been obtained using unitary simulations on three different $32^3\times64$ QCD+QED ensembles. Two of these ensembles possess exact U-spin symmetry ($m_d=m_s$), with one of these ensembles having $m_u$ tuned to approximate SU(3) symmetry as detailed in \cite{Horsley:2015vla}, and the remaining ensemble has three non-degenerate quarks. These ensembles are confined to a plane of constant average (bare) quark mass, $\bar{m} = (m_u+m_d+m_s)/3 = m_0 = constant$, and the quark hopping parameters used for each ensemble are given in Table~\ref{tab:DClattices}. The gauge actions used are the tree-level Symanzik improved SU(3) gauge action and the noncompact U(1) QED gauge action (further details in \cite{CSSM:2019jmq, Horsley:2015vla,Horsley:2015eaa}). The fermions are described by an $\mathcal{O}(a)$-improved stout link non-perturbative clover (SLiNC) action \cite{Cundy:2009yy}. The couplings used and resultant lattice spacing are 
\begin{equation}
\beta_{\textrm{QCD}} = 5.5, \quad \beta_{\textrm{QED}}=0.8, \quad a = 0.068(2) \, \textrm{fm},
\end{equation}
which corresponds to a QED coupling $\alpha_\textrm{QED} \simeq 0.1$, roughly $10$ times larger than the physical value. In our calculations of QCD+QED so far \cite{Horsley:2015eaa,CSSM:2019jmq} the strategy has been to simulate at an artificial coupling $\alpha_\textrm{QED} \simeq 0.1$ and then interpolate between this point and pure QCD to the physical fine structure constant $\alpha_\textrm{QED} = 1/137$. This value was chosen so that electromagnetic effects can be easily seen, but is still small enough that we can expect them to scale linearly in $\alpha_\textrm{QED}$.

\begin{table}
	\centering
	\begin{tabular}{| p{.35cm} | p{1.5cm} p{1.6cm} p{1.3cm} | p{1.8cm} | }
		\hline
		\multicolumn{5}{|c|}{Lattice ensemble parameters} \\
		\hline
		\# & $\kappa_u$ & $\kappa_d$ & $\kappa_s$ & $M_{\pi^+}$ (MeV) \\
		\hline
		1 & 0.124362 & 0.121713 & 0.121713 & 430(4) \\
		\hline
		2 & 0.124440 & 0.121676 & 0.121676 & 424(3) \\
		\hline
		3 & 0.124508 & 0.121821 & 0.121466 & 341(4) \\
		\hline
	\end{tabular}
	\caption{The number labels, $\kappa$-values and charged pion masses of the three $V_4=32^3\times 64$ ensembles employed for the determination of decay constants.} \label{tab:DClattices}
\end{table}

\subsection{Flavour-breaking expansions \label{sec:fbexpansions}}

The lattice simulations used in this study possess larger-than-physical quark masses. The mass parameters have been chosen near an SU(3) symmetric point, where the strong quark-mass symmetry breaking effects are relatively small, and hence comparable to the anticipated electromagnetic effects (at the large $\alpha_{\rm QED}$ of our calculations.  
We therefore seek to simultaneously describe the symmetry breaking effects from 
the quark masses and charges within the flavour-breaking expansion framework introduced in Ref.~\cite{Bietenholz:2011qq}.

It was previously shown in \cite{QCDSF-UKQCD:2016rau} that weak decay constants, defined through the couplings of pure octet currents to the corresponding octet states, $\langle 0 | A^a_\mu | n \rangle$, exhibit the same quark-flavour symmetries as the masses-squared of those states, and so for the $K^0$ we may simply co-opt the mass expansion, including QED effects, given in \cite{Horsley:2015vla} to write the leading-order expression
\begin{equation}
	F_{K^0} = F_0 \, + \, H^\textrm{EM}_0(e_u^2+e_d^2+e_s^2) \, + \, G(\delta m_d+\delta m_s). \label{eqn:k0Fexp}
\end{equation}
The mass parameters $\delta m_f=m_f - m_0$ denote the deviation of the mass of the $f$ quark flavour from the SU(3) symmetric point.

The flavour-neutral decay constant extrapolations are less straightforwardly deduced. As in the above case, the expansions relevant to decay constants defined through the couplings of quark-flavour currents to un-physical quark-flavour states, $F_{f f^\prime} = \langle 0 | \mathcal{A}_{f} | f^\prime \rangle / M_{f^\prime}$, may be deduced from the mass-squared expansions of those states, which have been given in \cite{CSSMQCDSFUKQCD:2021rvs}, to write
\begin{equation*}
	\begin{bmatrix} 
		F_{uu} & F_{ud} & F_{us} \\
		F_{ud} & F_{dd} & F_{ds} \\
		F_{us} & F_{ds} & F_{ss} \\
	\end{bmatrix} = \left[F_0  \, +  \, H^\textrm{EM}_0(e_u^2+e_d^2+e_s^2)\right]\cdot\mathbb{I}
\end{equation*}
\begin{equation*}
	  \, + E \begin{bmatrix} 
		1 & 1 & 1 \\
		1 & 1 & 1 \\
		1 & 1 & 1 \\
	\end{bmatrix}  \, +  \, h_0\begin{bmatrix} 
		2\delta m_u & \delta m_u+\delta m_d & \delta m_u+\delta m_s \\
		\delta m_u+\delta m_d & 2\delta m_d & \delta m_d+\delta m_s \\
		\delta m_u+\delta m_s & \delta m_d+\delta m_s & 2\delta m_s \\
	\end{bmatrix} 
\end{equation*}
\begin{equation*}
	+ \, 2H_1^\textrm{EM}\begin{bmatrix} 
		e_u^2 & 0 & 0 \\
		0 & e_d^2 & 0 \\
		0 & 0 & e_s^2 \\
	\end{bmatrix} \, + \, 2G \begin{bmatrix} 
		\delta m_u & 0 & 0 \\
		0 & \delta m_d & 0 \\
		0 & 0 & \delta m_s \\
	\end{bmatrix} 
\end{equation*}
\begin{equation}
     \, +  \, g_1^\textrm{EM}\begin{bmatrix} 
		e_u^2 & e_u e_d & e_u e_s \\
		e_u e_d & e_d^2 & e_d e_s \\
		e_u e_s & e_d e_s & e_s^2 \\
	\end{bmatrix}. \label{eqn:FNDCexp}
\end{equation}
To render the above expression applicable to the FN decay constants we calculate herein, defined in Equation~\eqref{eqn:QFBdecayconstdef} with respect to physical states, we may `rotate' the quark-flavour states through the second index $f'$ using the flavour compositions which we have previously determined in \cite{CSSMQCDSFUKQCD:2021rvs},
\begin{equation}
    c_n^f \equiv \frac{\langle 0 | {\mathcal{P}}_{f} | n \rangle}{\sqrt{\sum_{f^\prime = u,d,s} \langle 0 | {\mathcal{P}}_{f^\prime} | n \rangle }},
\end{equation}
which at large Euclidean times satisfy
\begin{equation}
	\langle 0 | \, \sum^3_{f=1} \, c_n^f \, \mathcal{P}_f \, | m \rangle \propto \delta_{nm}, \quad n,m  = \pi^0,\eta,\eta^\prime. \label{eqn:pietaevecdelta}
\end{equation}
The FN PS decay constants we determine on the lattice are hence fitted to the LO expression
\begin{equation}
	\begin{bmatrix} 
		F^u_{\pi^0} & F^u_{\eta} & F^u_{\eta^\prime} \\
		F^d_{\pi^0} & F^d_{\eta} & F^d_{\eta^\prime} \\
		F^s_{\pi^0} & F^s_{\eta} & F^s_{\eta^\prime} \\
	\end{bmatrix} =
	\begin{bmatrix} 
		F_{uu} & F_{ud} & F_{us} \\
		F_{ud} & F_{dd} & F_{ds} \\
		F_{us} & F_{ds} & F_{ss} \\
	\end{bmatrix}
	\begin{bmatrix} 
		c^u_{\pi^0} & c^u_{\eta} & c^u_{\eta^\prime} \\
		c^d_{\pi^0} & c^d_{\eta} & c^d_{\eta^\prime} \\
		c^s_{\pi^0} & c^s_{\eta} & c^s_{\eta^\prime} \\
	\end{bmatrix}, \label{eqn:DCfitfunc}
\end{equation}
where the $c_n^f$ shown above are the mass/charge parametrizations introduced and fitted to results from a variety of lattices in \cite{CSSMQCDSFUKQCD:2021rvs}.

\subsection{RI$^\prime$-MOM results \label{sec:rimomresults}}

The RI$^\prime$-MOM renormalization Z-factors are determined on Ensemble~1. For the matrix of FN axial-vector Z-factors, which have disconnected contributions, the quark-loop terms $\textrm{Tr}\left[ S_f(z;z)\gamma_5\gamma_3 \right]$ are calculated using colour- and spin-diluted $\mathbb{Z}_2$ volume-sources with $\mathcal{O}(100)$ independent stochastic sources on each gauge configuration. 

\begin{figure*}
	\centering
	\includegraphics[width=0.9\textwidth]{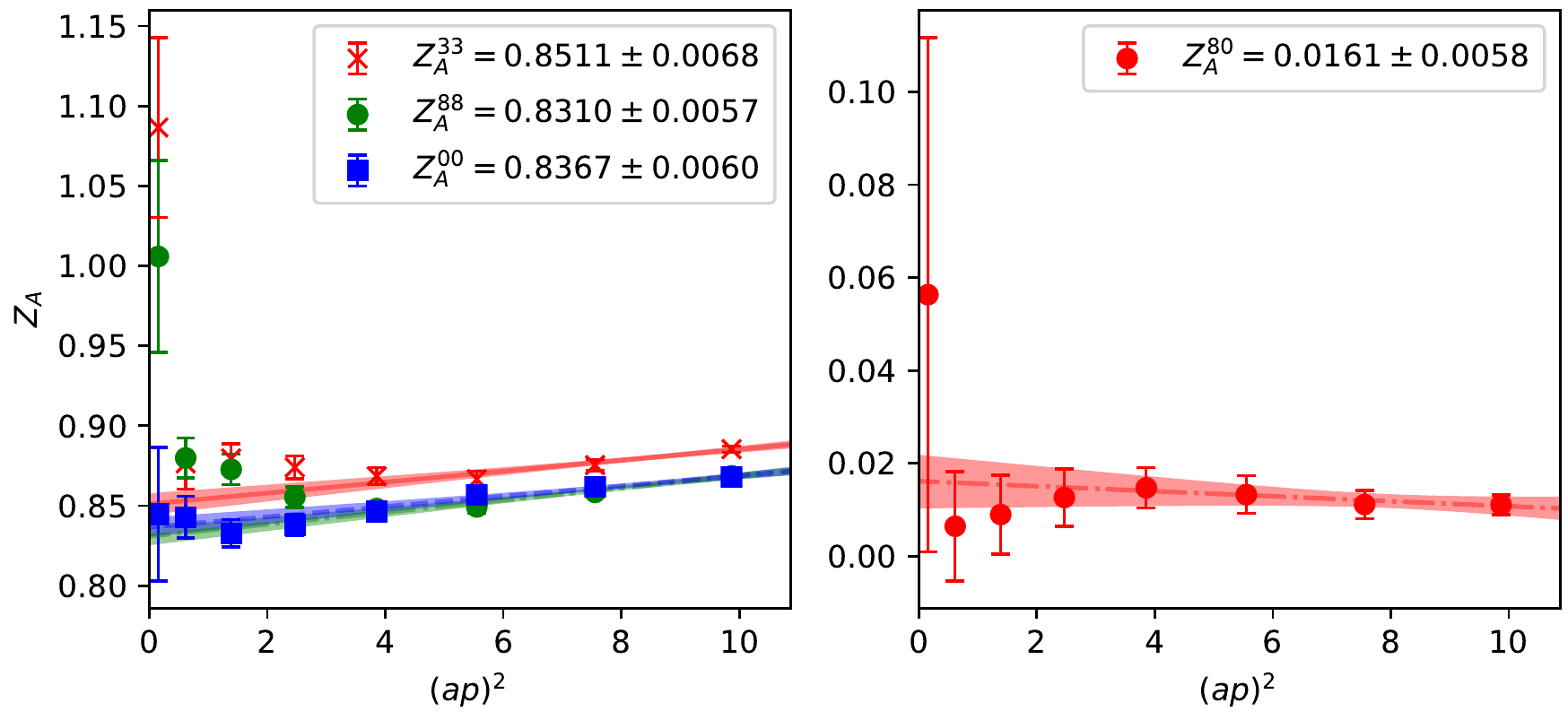}
	\caption{The distinct flavour-diagonal (LHS) and off-diagonal (RHS) elements of the axial-vector Z-factor matrix given in Equation~\eqref{eqn:matrixZs}. The flavour-diagonal factors have a large connected contribution and are hence well resolved, whilst the off-diagonals are purely disconnected (see Equation~\eqref{eqn:3ptGreenfunc}).} \label{fig:convsfullZfacs}
\end{figure*}

For our determinations of the off-diagonal, purely disconnected elements of the Z-factor matrix given in Equation~\eqref{eqn:matrixZs}, we do not find any statistical difference from zero. However, transforming this matrix into its U-spin equivalent, where the $d$ and $s$ quarks play the role usually played by the $u$ and $d$ quarks in the isospin basis, via the rotations:
\mathleft
\begin{equation*}
    \begin{bmatrix} 
		Z^{33}_A & Z^{38}_A & Z^{30}_A \\
		Z^{83}_A & Z^{88}_A & Z^{80}_A \\
		Z^{03}_A & Z^{08}_A & Z^{00}_A \\
	\end{bmatrix} =
\end{equation*}
\mathcenter
\begin{equation}
    	\begin{bmatrix} 
		0 & \frac{1}{\sqrt{2}} & \frac{-1}{\sqrt{2}} \\
	    \frac{-2}{\sqrt{6}} & \frac{1}{\sqrt{6}} & \frac{1}{\sqrt{6}} \\
		\frac{1}{\sqrt{3}} & \frac{1}{\sqrt{3}} & \frac{1}{\sqrt{3}} \\
	\end{bmatrix}
		\begin{bmatrix} 
		Z^{uu}_A & Z^{ud}_A & Z^{us}_A \\
		Z^{du}_A & Z^{dd}_A & Z^{ds}_A \\
		Z^{su}_A & Z^{sd}_A & Z^{ss}_A \\
	\end{bmatrix} 
    	\begin{bmatrix} 
		0 & \frac{-2}{\sqrt{6}} & \frac{1}{\sqrt{3}} \\
	    \frac{1}{\sqrt{2}} & \frac{1}{\sqrt{6}} & \frac{1}{\sqrt{3}} \\
		\frac{-1}{\sqrt{2}} & \frac{1}{\sqrt{6}} & \frac{1}{\sqrt{3}} \\
	\end{bmatrix},
\end{equation}
we can isolate the non-trivial mixing contribution as $Z^{80}_A$, which arises at leading order from a purely connected, QED vertex. Put differently, the lattice signal for $Z^{80}_A$ is dominated by the difference of the connected contributions coming from $\bar{u}u$- and $\bar{d}d$/$\bar{s}s$-type operators due to QED. We find this U-spin matrix to be symmetric within statistical errors, which is non-trivial since the indices label the external states and operator insertion respectively in the renormalization scheme (see Equation~\eqref{eqn:3ptGreenfunc}). In Figure~\ref{fig:convsfullZfacs} we present the diagonal elements of the U-spin Z-matrix (LHS), as well as those unique off-diagonal elements (RHS) which are not statistically consistent with zero. In each case the linear extrapolation to $(ap)^2=0$ is performed using only the four largest values of momenta.

\begin{figure}
	\centering
	\includegraphics[width=0.48\textwidth]{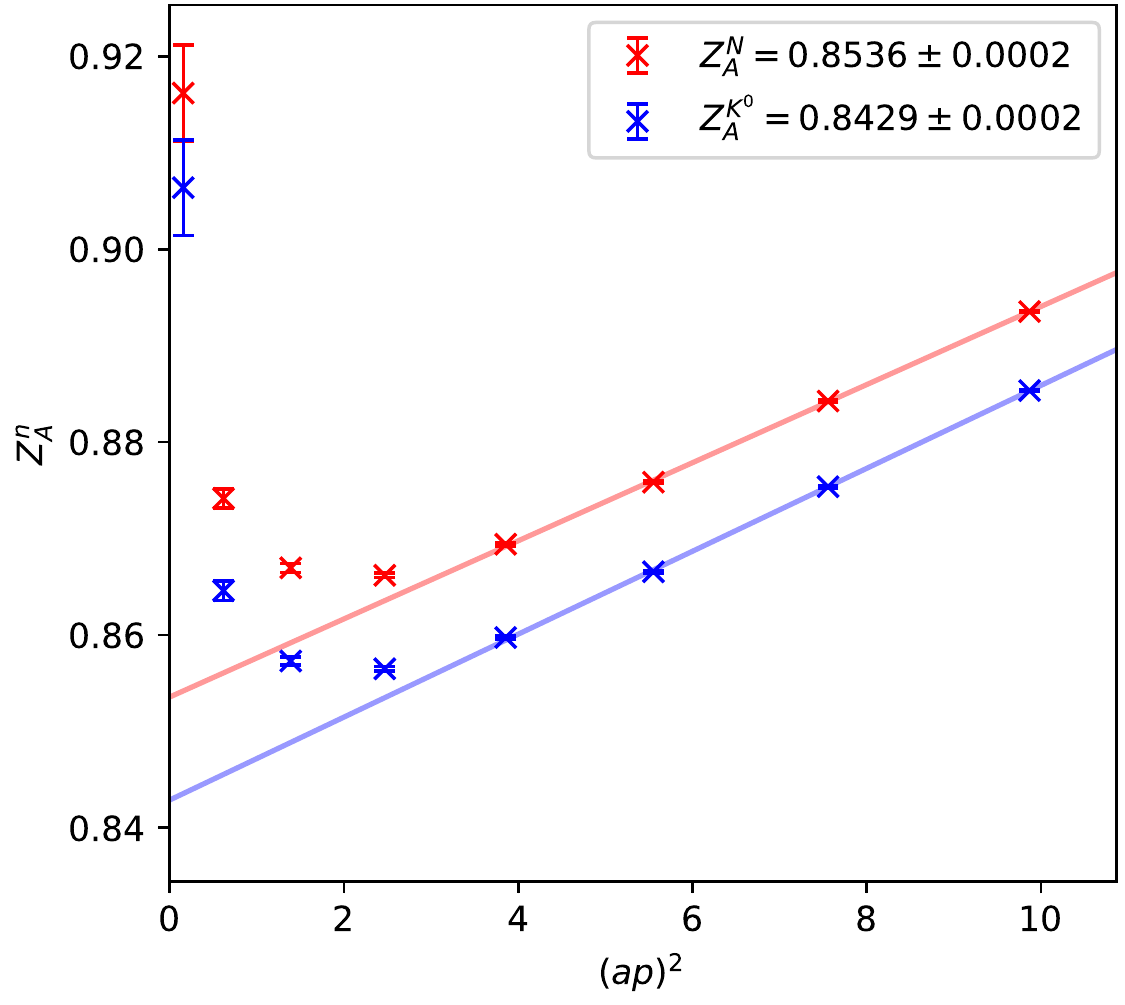}
	\caption{The axial-vector Z-factor applicable to the $K^0$ decay constant, which receives only connected contributions, and an additional Z-factor $Z_{A}^{N}$ calculated for a fictitious neutral-quark operator which is shown to give a sense of the valence charge dependence (further details given in Section~\ref{sec:rimomresults}).} \label{fig:outerringZfacs}
\end{figure}

In Figure~\ref{fig:outerringZfacs} we present the Z-factor required for renormalization of the $K^0$ decay constants, as well as a fictitious neutral, connected-only Z-factor for comparison, which is calculated using partially-quenched quarks with zero electric charge and hopping parameter $\kappa_n=0.1208142$. The mass of this neutral quark has been tuned so that $M^2_{n\bar{n}}=M^2_{u\bar{u}}=M^2_{d\bar{d}}$, without regard for disconnected contributions, in keeping with the Dashen scheme outlined in Ref.~\cite{Horsley:2015vla}. The numbers appearing in the legend in each case correspond to the $(ap)^2\to 0$ limit, where the fit is performed on only the four largest momentum values for which the expected linear $(ap)^2$ scaling is observed.

\subsection{Decay constant results \label{sec:dcresults}}

As detailed in Ref.~\cite{CSSMQCDSFUKQCD:2021rvs} we calculate all connected contributions to our correlation functions using the one-end trick \cite{Alexandrou:2008ru}, and again utilise three independent $\mathbb{Z}_2$-noise sources per gauge field configuration in calculating both disconnected and connected diagrams. As per Equations~\eqref{eqn:rawlatticeFNdecayconst} and \eqref{eqn:rawpi+decayconst} we also calculate the meson masses in the determinations of the decay constants, and for each of the three ensembles 1--3 they are given in Table~\ref{tab:pieta32masses}, Appendix~\ref{appendix}.

\begin{figure}
	\centering
	\includegraphics[width=0.48\textwidth]{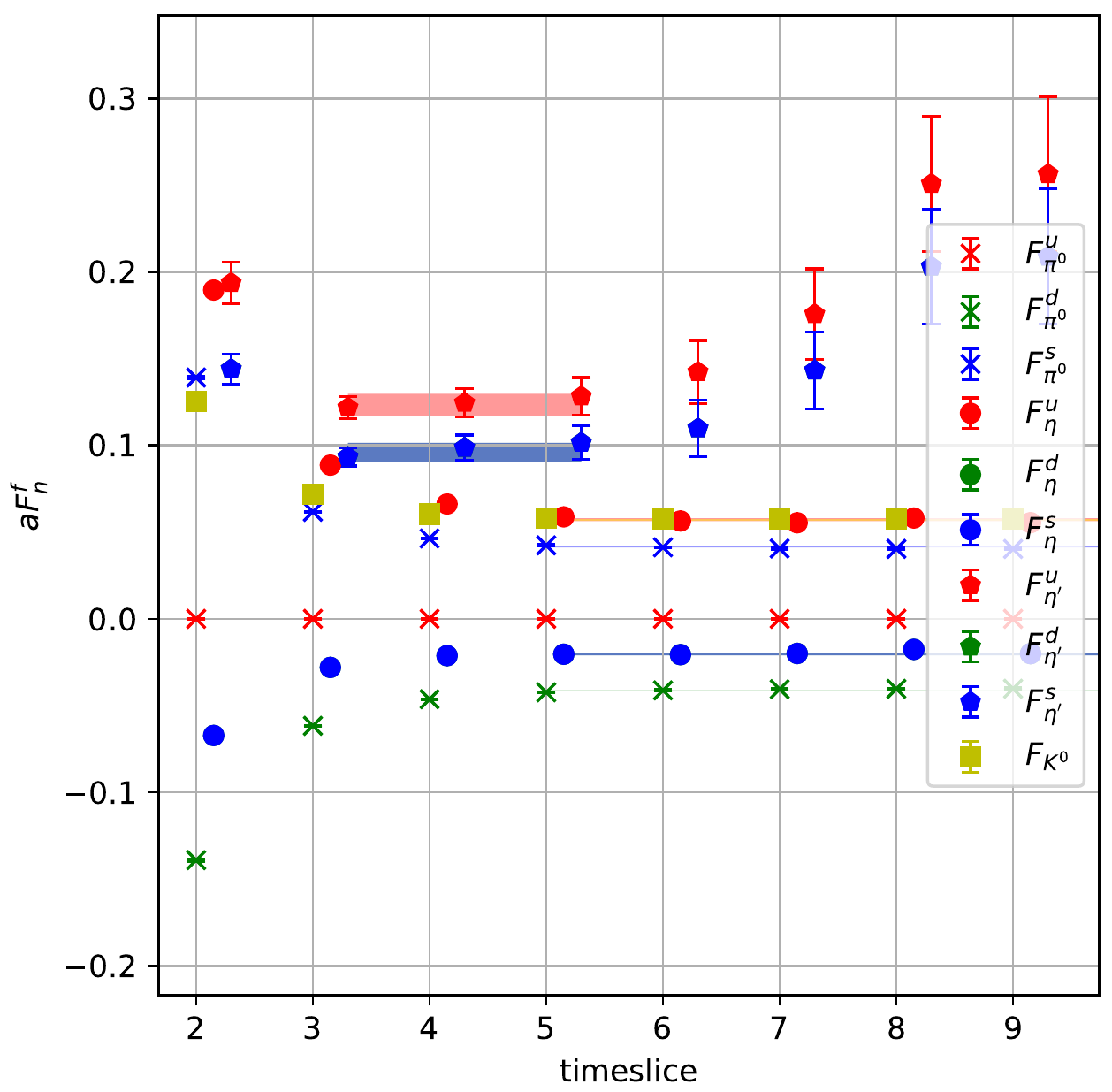}
	\caption{An example of the decay constant signals as obtained from Equation~\eqref{eqn:rawlatticeFNdecayconst} (FN decay constants), and Equation~\eqref{eqn:rawpi+decayconst} ($F_{K^0}$; yellow squares), from Ensemble~1 with $t_0=4$. Note that the down and strange quarks here are degenerate and so the signals corresponding to those flavours have the same magnitudes. Also pictured are constant fits to the plateau regions in each case. See Section~\ref{sec:dcresults} for more discussion of the $\eta^\prime$ signal.} \label{fig:ensemble1DCs}
\end{figure}

In Figure~\ref{fig:ensemble1DCs} we give an example of the signals obtained for the (non-renormalized) decay constants on our Ensemble~1. The eigenvectors appearing in Equation~\eqref{eqn:rawlatticeFNdecayconst} are calculated using the GEVP with $t_0=4$ and $\delta t =1$, as the GEVP becomes unstable at larger $t_0$. As can be seen, the $\eta^\prime$ signal quality tends to degrade at relatively small values of Euclidean time, forcing us to fit a plateau earlier than for the octet-dominant species. Moreover, the singlet-dominant (i.e. $\eta'$) decay constant appears to increase at larger times, which is a result of the corresponding eigenvector failing to diagonalize the correlation matrix at times much greater than $t_0$, which subsequently results in the presence of exponential growth in Equation~\eqref{eqn:rawlatticeFNdecayconst}. It is hence reasonable to assume that our extracted values of the $\eta^\prime$ decay constants are poorly determined.
The renormalized FN and $K^0$ decay constants are given in Table~\ref{tab:pieta32decayconsts}, Appendix~\ref{appendix}. 

\begin{figure*}
	\centering	\includegraphics[width=0.85\textwidth]{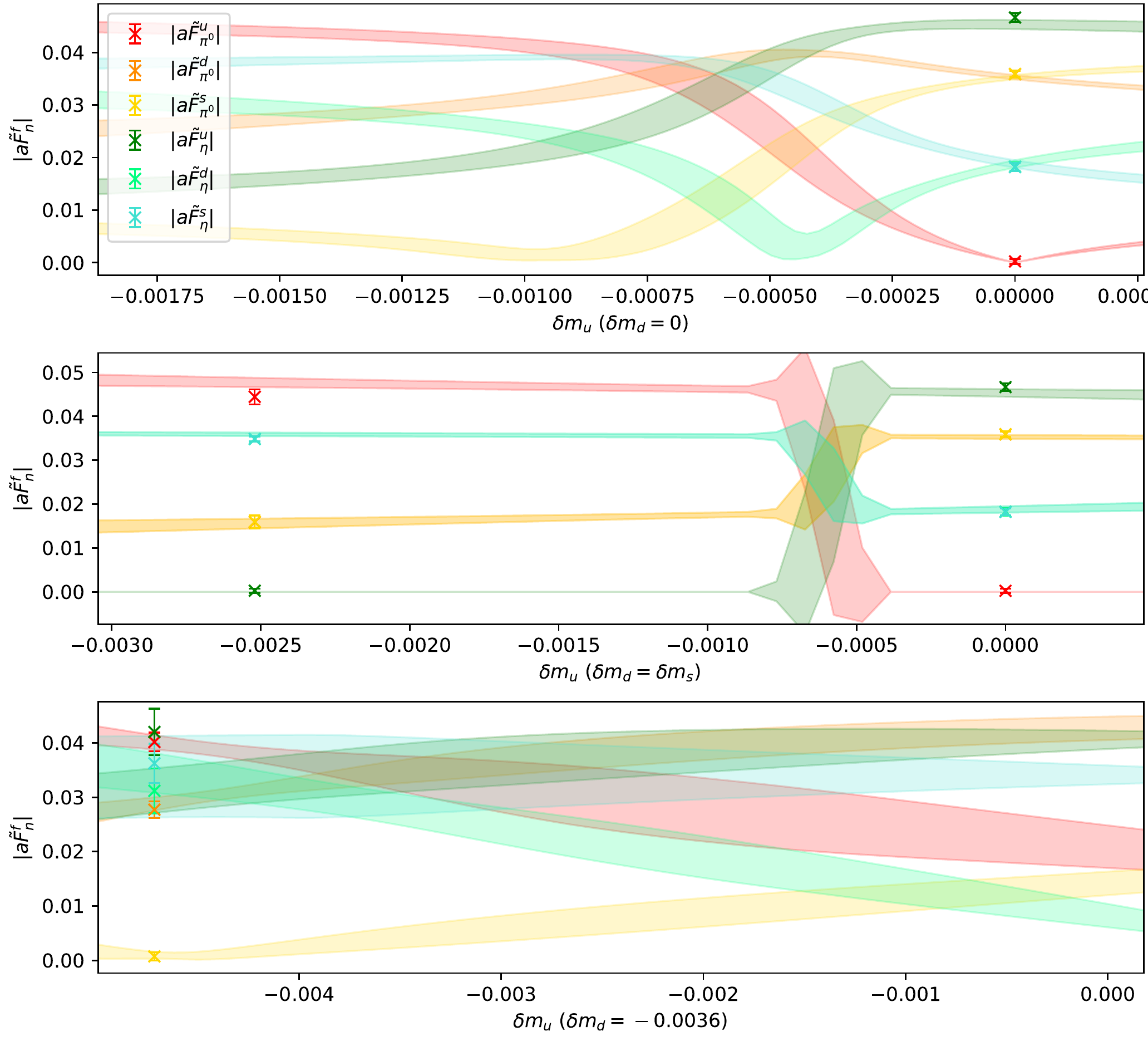}
	\caption{The RI$^\prime$-MOM scheme $\pi^0$ and $\eta$ decay constant magnitudes calculated on each of our ensembles 1--3, along with the global fit described in Section~\ref{sec:fbexpansions}. The top panel depicts the results obtained using Ensemble~1 along with the fit on the trajectory $\delta m_d =0$. The centre panel depicts ensembles 1--2 (right-to-left) along the exact U-spin trajectory. The bottom panel illustrates the fit around, and results of, Ensemble~3 which possesses the greatest broken flavour-symmetry of our ensembles.} \label{fig:FNDCsfit}
\end{figure*}

The fitting of the quark-mass extrapolation scheme for the decay constants, described in Section~\ref{sec:fbexpansions}, is performed on the lattice decay constants with the RI$^\prime$-MOM Z-factors applied according to Equation~\eqref{eqn:FNrenormoperator},
\begin{equation}
	\tilde{F}^f_n \equiv \sum_{f'} Z_A^{f f'} F^{f'}_n.
\end{equation}
We have tried performing the decay constant fit with all neutral PS species considered, but seemingly because of our difficulty with determinations of the $\eta^\prime$ decay constants, as discussed above, we are not able to achieve a fit with acceptable quality. Instead, we presently perform the fit to only the $\pi^0$, $\eta$ and $K^0$ decay constants, which gives $\chi^2/\textrm{dof}=1.6$. The $\pi^0$ and $\eta$ decay constant fits are presented in Figure~\ref{fig:FNDCsfit}. 

The top panel of Figure~\ref{fig:FNDCsfit} displays the decay constant fit on a quark-mass trajectory with $\delta m_d = 0$, on which our Ensemble~1 results are presented. In the second panel of Figure~\ref{fig:FNDCsfit} we present the fit along the U-spin trajectory ($\delta m_d = \delta m_s$), where one of the $\pi^0$ or $\eta$ possesses the flavour-composition of a U-spin pion: $\pi^U_3 = (\bar{d}d-\bar{s}s)/\sqrt{2}$. Also, quantities relating to down and strange quarks are necessarily equal along this trajectory.
The final (bottom) panel of Figure~\ref{fig:FNDCsfit} gives the fit along the constant-$\delta m_d$ trajectory which our Ensemble~3 occupies, and which possesses the lightest up and down quarks of any of our ensembles 1--3. It is clear that with our limited set of ensembles, the mass extrapolation becomes poorly constrained at such large values of SU(3)-flavour breaking as are present in this panel.

\begin{table}
	\centering
	\begin{tabular}{| p{0.5cm} | p{1.7cm} p{1.7cm} p{1.5cm} |}
		\hline
		\multicolumn{4}{|c|}{Physical-point decay constants (MeV)} \\
		\hline
		 & $l={\pi_3}$ & $l={\eta_8}$ & $l={\eta_1}$ \\
		\hline
		$\tilde{F}^l_{\pi^0}$ & 133(41) & -19(44) & 33(32) \\
		\hline
		$\tilde{F}^l_{\eta}$ & -10(24) & 143(54) & 85(27) \\
		\hline
  
	\end{tabular}
	\caption{The values of the $\pi^0$ and $\eta$ decay constant fits at the physical-point quark masses in the RI$^\prime$-MOM scheme, at the approximate scale $\mu\approx a^{-1}$. The corresponding physical-point value obtained for $F_{K^0}$ is 146(3) MeV. The quoted uncertainties are statistical only.} \label{tab:physFNDCs}
\end{table}

Whilst our current inability to determine and fit the $\eta^\prime$ decay constants with confidence means that we cannot presently make an estimate of the full FN PS sector decay constants at the physical point, it is still of interest to make an assessment of the remaining decay constants, and to that end we present the values of our fits at the physical point determined for $32^3\times 64$ volume \cite{Horsley:2015vla},
\begin{equation}
	a\delta m^\star_u = -0.00834(8), \quad a\delta m^\star_d = -0.00776(7).
\end{equation}
Since isospin is a good approximate symmetry at the physical point, it is most intuitive to present the physical point determination in the (isospin) octet-singlet basis, and so with the RI$^\prime$-MOM renormalization factors considered we have
\begin{equation}
	\tilde{F}^{\pi_3}_n = \frac{1}{\sqrt{2}}\left( \tilde{F}^u_n - \tilde{F}^d_n \right), \quad \tilde{F}^{\eta_8}_n = \frac{1}{\sqrt{6}}\left( \tilde{F}^u_n + \tilde{F}^d_n -2\tilde{F}^s_n \right),
\end{equation}
\begin{equation}
	\tilde{F}^{\eta_1}_n = \frac{1}{\sqrt{3}}\left( \tilde{F}^u_n + \tilde{F}^d_n + \tilde{F}^s_n \right),
\end{equation}
for each $n\in [\pi^0,\eta]$. Moreover, we have scaled all expansion parameters which originate with QED by the factor $\alpha^\star_\textrm{QED}/\alpha_\textrm{QED}=0.073$, to approximately correct for our unphysical EM coupling. Unfortunately, since we have no simulations for varying quark charges, we cannot properly constrain the shared $H^\textrm{EM}_0$ term, and so some residual larger-than-physical QED effect will remain. 

The $\pi^0$and $\eta$ physical-point results are given in Table~\ref{tab:physFNDCs}.
As we found in Ref.~\cite{CSSMQCDSFUKQCD:2021rvs} for the state compositions, which inform the decay constants presented here via Equation~\eqref{eqn:DCfitfunc}, our results for the flavour-diagonal decay constants are consistent with isospin symmetry since we do not resolve non-zero $\tilde{F}^{\eta_8}_{\pi^0}$ or $\tilde{F}^{\eta_1}_{\pi^0}$. 
It should be noted that in the uncertainties 
quoted are statistical only, and a quantitative assessment of the uncertainties arising from discretization, finite volume and quark mass extrapolation are left for future work.

Considering now the $\eta$ decay constants, for comparison we take the results of the recent lattice study \cite{Bali:2021qem} at $\mu=2\, \textrm{GeV}$, distinguished as $\hat{F}^a_{\eta}$, which have central values $\hat{F}^{\eta_8}_{\eta} = 149 \, \textrm{MeV}$ and $\hat{F}^{\eta_1}_{\eta} = 16 \, \textrm{MeV}$. The associated uncertainties are all $\mathcal{O}(10\%)$, and we note that these results were obtained assuming isospin symmetry and without QED. Comparing these numbers with the relevant results of Table~\ref{tab:physFNDCs} we see excellent agreement for the $\eta_8$ component, whilst our fit favours a slightly larger singlet component than was found in \cite{Bali:2021qem}, although the uncertainty is large. 

\section{Conclusion}

In this study we have built on the theoretical and numerical machinery developed in our previous work \cite{CSSMQCDSFUKQCD:2021rvs} to extract the weak decay constants of the FN and $K^0$ PS mesons. We have also introduced and presented preliminary results for the extension to the RI$^\prime$-MOM renormalization scheme which facilitates a proper treatment of the renormalization of FN currents on the lattice.

We have deduced quark-mass and charge expansions for the neutral PS meson decay constants, finding that proper parametrization of the FN sector requires knowledge of the relevant underlying PS flavour-compositions. These parametrizations facilitate physical-point determinations of the neutral PS decay constants which, with improved precision and an improved signal for the $\eta^\prime$, may be used as numerical inputs for theoretical and phenomenological calculations regarding various (rare) leptonic decays which include PS mesons, e.g. $\pi^0/\eta/\eta^\prime\rightarrow e^+e^-$.

Through the results of this work we have demonstrated a process by which precision physical-point determinations of all neutral PS meson decay constants, including strong isospin-breaking and QED, should in principle be possible. To this end one must first improve upon the decay constant determination of the $\eta^\prime$, which is primarily limited by ones ability to precisely compute disconnected contributions to the relevant correlation functions, however for a precision study one would also desire infinite volume and continuum extrapolations, as well as a careful treatment of the scale dependence of the FN operators.

\begin{acknowledgments}
The numerical configuration generation (using the BQCD lattice QCD program \cite{Haar:2017ubh} with single quark flavours treated in the HMC by the tRHMC algorithm \cite{Haar:2018jjd}) and data analysis (using the Chroma software library \cite{Edwards:2004sx} and a GPU-accelerated mixed-precision conjugate gradient fermion matrix inverter through the COLA software \cite{Kamleh:2012sh}) was carried out on the DiRAC Blue Gene Q and Extreme Scaling (EPCC, Edinburgh, UK) and Data Intensive (Cambridge, UK) services, the GCS supercomputers JUQUEEN and JUWELS (NIC, Jülich, Germany) and resources provided by HLRN (The North-German Supercomputer Alliance), the NCI National Facility in Canberra, Australia (supported by the Australian Commonwealth Government) and the Phoenix HPC service (University of Adelaide). ZRK was supported by an Australian Government Research Training Program (RTP) Scholarship. 
RH was supported by STFC through grant ST/P000630/1.
WK was supported by Australian Research Council Grants DP19012215, DP210103706 and the Pawsey Centre for Extreme Scale Readiness.
HP was supported by DFG Grant No. PE 2792/2-1.
PELR was supported in part by the STFC under contract ST/G00062X/1.
GS was supported by DFG Grant No. SCHI 179/8-1.
RDY and JMZ were supported by the Australian Research Council grant DP190100297 and DP220103098.
We thank all funding agencies.
For the purpose of open access, the authors have applied a Creative Commons Attribution (CC BY) licence to any author accepted manuscript version arising from this submission.
\end{acknowledgments}

\appendix
\section{Lattice results\label{appendix}}

\begin{table*}
	\centering
	\begin{tabular}{| p{.4cm} p{1.4cm}  p{1.4cm}  p{1.6cm}  p{1.8cm}   p{1.4cm} |}
		\hline
		\multicolumn{6}{|c|}{Neutral PS meson masses (MeV)} \\
		\hline
		\# & $M_{\pi^0}$ & $M_{\eta}$ & $M_{\eta}$-$M_{\pi^0}$ & $M_{\eta^\prime}$ & $M_{K^0}$ \\
		\hline
		1 & 400(4) & 424(5) & 15(1) & 1134(179) & 400(4) \\
		\hline
		2 & 401(7) & 441(3) & 71(6) & 1239(69) & 441(3) \\
		\hline
		3 & 312(8) & 470(18) & 253(17) & 1097(98) & 486(2) \\
		\hline
		
	\end{tabular}
	\caption{The extracted neutral PS meson masses, as well as the mass-splitting $M_\eta-M_{\pi^0}$, for each of our three $32^3\times 64$ ensembles. \label{tab:pieta32masses}}
\end{table*}
\begin{table*}
	
		\centering
\begin{tabular}{| p{.4cm}  p{1.6cm}  p{1.6cm}  p{1.6cm}  p{1.6cm}  p{1.6cm} p{1.6cm}  p{1.6cm}  p{1.6cm}  p{1.6cm} | p{1.2cm} |}
	\hline
	\multicolumn{11}{|c|}{Neutral PS decay constants (MeV)} \\
	\hline
	\# & $\tilde{F}^u_{\pi^0}$ & $ \tilde{F}^d_{\pi^0}$ & $\tilde{F}^s_{\pi^0}$ & $\tilde{F}^u_{\eta}$ & $\tilde{F}^d_{\eta}$ & $\tilde{F}^s_{\eta}$ & $\tilde{F}^u_{\eta^\prime}$ & $\tilde{F}^d_{\eta^\prime}$ & $\tilde{F}^s_{\eta^\prime}$ & $\tilde{F}_{K^0}$ \\
	\hline
	1 & 1(1) & 101(2) & 101(2) & 133(2) & 50(2) & 50(2) & 288(14) & 233(13) & 233(13) & 140(1) \\
	\hline
	2 & 142(4) & 33(3) & 33(3) & 1(1) & 100(2) & 100(2) & 242(11) & 213(10) & 213(10) & 141(1) \\
	\hline
	3 & 125(10) & 80(9) & 8(4) & 104(22) & 107(15) & 106(8) & 203(34) & 205(35) & 244(19) & 145(1) \\
	\hline
	
\end{tabular}
\caption{The renormalized quark-flavour basis decay constants of the physical FN states, and the renormalized $K^0$ decay constants, on each of our $32^3\times 64$ ensembles in the RI$^\prime$-MOM scheme. \label{tab:pieta32decayconsts}}
\end{table*}

\bibliography{npsmixing}

\end{document}